\documentclass[12pt,a4paper]{conference}

\usepackage{fancyhdr}
\usepackage{graphicx,amsmath,amssymb,cite}
\usepackage{multind}
\makeindex{author} \makeindex{subject}

\newcommand{\beq}{\begin{equation}}
\newcommand{\eeq}[1]{\label{#1}\end{equation}}
\newcommand{\eeqn}{\end{equation}}

\newcommand{\beqa}{\begin{eqnarray}}
\newcommand{\eeqa}[1]{\label{#1}\end{eqnarray}}
\newcommand{\eeqan}{\end{eqnarray}}

\let\bar=\overbar

\newcommand{\Dslash}{\not{\hbox{\kern-4pt $D$}}}
\newcommand{\dslash}{\not{\hbox{\kern-2pt $\del$}}}

\newcommand{\msb}{{\bar{\ssstyle M \kern -1pt S}}}

\begin{document}

\Chapter{Tetraquark spectroscopy}
           {Tetraquark spectroscopy}{E. Santopinto \it{et al.}}
\vspace{-6 cm}\includegraphics[width=6 cm]{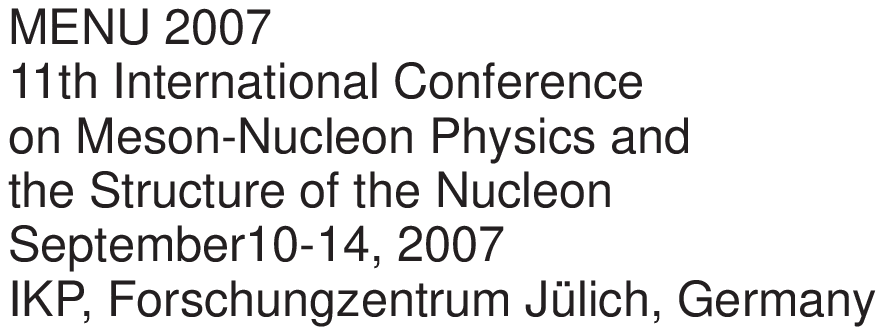}
\vspace{4 cm}

\addcontentsline{toc}{chapter}{{\it E. Santopinto}} \label{authorStart}

\begin{raggedright}

{\it E. Santopinto }\index{author}{Santopinto, E.}\\
Department of Physics\\
INFN and University of Genova\\
via Dodecaneso 33, Genova, Italy 16146
\bigskip\bigskip


{\it G. Galat\`a }\index{author}{Galat\`a, G.}\\
Department of Physics\\
University of Genova\\
via Dodecaneso 33, Genova, Italy 16146
\bigskip\bigskip

\end{raggedright}

\begin{center}
\textbf{Abstract}
\end{center}
A complete classification of tetraquark states in terms of the spin-flavor, color and spatial degrees of freedom was constructed. 
The permutational symmetry properties of both the spin-flavor and orbital parts of the 
quark-quark and antiquark-antiquark  subsystems are discussed.
This complete classification is general and model-independent, and is 
useful both for
model-builders and experimentalists. 
The total wave functions are also explicitly constructed in the hypothesis of ideal mixing;
this basis for tetraquark states will enable the eigenvalue problem to be solved for a definite
dynamical model. This is also valid for diquark-antidiquark models, for which the basis is a
subset of the one we have constructed.
An evaluation of the tetraquark spectrum is obtained from the Iachello mass formula for normal mesons, here generalized to tetraquark systems. This mass formula is a generalizazion of the Gell-Mann Okubo mass formula, whose coefficients have been upgraded by means of the latest PDG data.
 The ground state tetraquark nonet was identified with
$f_{0}(600)$, $\kappa(800)$, $f_{0}(980)$, $a_{0}(980)$.  The mass splittings predicted
 by this mass formula are compared to the KLOE, Fermilab E791 and BES experimental data. The diquark-antidiquark limit was also studied.

\section{\label{sec:introduzione}Introduction}

Light meson spectroscopy, in particular the nature of the scalar nonet, is still an open problem. Recently the KLOE, E791 and BES collaborations have provided evidence of the low mass resonances $f_{0}(600)$ \cite{Aloisio:2002bt} \cite{Aitala:2000xu} \cite{Ablikim:2005ni} and $\kappa(800)$ \cite{Aitala:2000xu} \cite{Ablikim:2005ni}. The quark-antiquark assignment to P-waves \cite{Tornqvist:1995kr} has never
 worked for the lowest lying scalar mesons, $f_{0}(980)$, $a_{0}(980)$, $\kappa(800)$ and 
$f_0(600)$ \cite{Jaffe:1976ig}. Maiani {\it et al.} in Ref.\cite{Maiani:2004uc} have suggested that these mesons could be described as a tetraquark nonet, in particular as a diquark-antidiquark system.  
In the traditional quark-antiquark scheme, the $f_{0}(980)$ is made up of non-strange quarks \cite{Tornqvist:1995kr} and so it is difficult to explain both its higher mass respect to the other components of the nonet and its decay properties (see Refs.\cite{Jaffe:1976ig} \cite{Maiani:2004uc}).  Already in the seventies Jaffe \cite{Jaffe:1976ig} suggested the tetraquark structure of the scalar nonet and proposed a four quark bag model. Other identifications, in particular 
as quasimolecular-states in Ref.\cite{Weinstein:1982gc} and as dynamically generated resonances in Ref.\cite{Oset}, have been proposed (for a complete review see Refs.\cite{Amsler:2004ps,Close:2002zu,PDG} and references therein).

We present here a complete classification scheme of the two quark-two antiquark states in terms of SU(6)$_{~sf}$ from Ref.\cite{Santopinto:2006my}, as well as
an evaluation of the tetraquark spectrum for the lowest scalar meson nonet, obtained from a generalization, to the tetraquark case, of the Iachello mass formula
for normal mesons published in Ref.\cite{Iachello:1991fj}. 

Since the classification of the states is general, it is valid whichever dynamical model for tetraquarks is chosen. As an application, in section \ref{sec:diqantidiq} we develop a simple diquark-antidiquark model with no spatial excitations inside diquarks.  In this case the states are a subset of the general case. 

\section{\label{sec:classificazionestati}The classification of tetraquark states}

In the construction of the classification scheme we shall make use of symmetry principles without, for the moment, introducing any explicit dynamical model.
We are constrained by two conditions: the tetraquark wave functions should be a colour singlet, as all physical states, and the tetraquarks states must be antisymmetric for the exchange of the two quarks and the two antiquarks.

First we begin with the internal (color, flavor and spin) degrees of freedom.
The allowed SU(3)$_{f}$ representations for the $qq\bar q\bar q$ system are obtained by means of the product $[3]\otimes [ 3]\otimes [\bar 3]\otimes [\bar 3] =[1]\oplus [8]\oplus [1]\oplus [8]\oplus [27]\oplus [8]\oplus [8]\oplus [10]\oplus [\overline{10}]$. The allowed isospin values are $I=0,\frac{1}{2},1,\frac{3}{2},2$ , while the hypercharge values are $Y=0,\pm 1,\pm 2$. The values $I=\frac{3}{2},2$ and $Y=\pm 2$ are exotic, which means that they are forbidden for the $q\bar q$ mesons.
The allowed SU(2)$_{s}$ representations are obtained by means of the product $ [2]\otimes [2]\otimes [2]\otimes [2] =[1]\oplus [3]\oplus [1]\oplus [3]\oplus [3]\oplus [5]$. 
The tetraquarks can have an exotic spin $S=2$, value forbidden for normal $q\bar q$ mesons.
The SU(6)$_{sf}$-spin-flavour classification is obtained
 by $[6]\otimes [ 6]\otimes [\bar 6]\otimes [\bar 6]=[1]\oplus [35]\oplus [405]\oplus [1]\oplus [35]\oplus [189]\oplus [35]\oplus [280]\oplus [35]\oplus [\overline{280}]$. In Appendices A and B of Ref.\cite{Santopinto:2006my} all the flavor and spin states in the $qq\bar q \bar q$ configuration are explicitly written in terms of the single quark and antiquark states. The flavor states are also written in the ideal mixing hypothesys, i.e. as a superposition of the SU(3)-symmetrical states in such a way to have defined strange quark and antiquark numbers. The ideal mixing is essentially a consequence of the OZI rule and, while it has not been proved yet, it is used by all the authors  working on $q\bar q$ mesons and tetraquarks.
 
We can now describe the spatial degrees of freedom. 
The tetraquark is a system made up of four objects. Thus, we have to define three relative coordinates that we choose as in Ref.\cite{Estrada}: a relative coordinate between the two quarks, another between the two antiquarks and the third relative coordinate between the centers of mass of the two $q$ and the two $\bar q$. We associate to each coordinate an orbital angular momentum, $L_{13}$, $L_{24}$ and $L_{12-34}$ respectively.  
We obtain the total angular momentum $J$ by combining the four different spins and the three orbital angular momenta.
The parity for a tetraquark system is the product of the intrinsic parities of the quarks (+) and the antiquarks (-) times the factors coming from the spherical harmonics \cite{Estrada}. The result is
$P=P_{q}P_{q}P_{\bar q}P_{\bar q}(-1)^{L_{13}}(-1)^{L_{24}}(-1)^{L_{12-34}}=(-1)^{L_{13}+L_{24}+L_{12-34}}$.
Using these coordinates, the charge conjugation eigenvalues can be calculated by following the same steps as in the $q\bar q$ case, considering a tetraquark as a $Q\bar Q$ meson, where $Q$ represents the couple of quarks and $\bar Q$ the couple of antiquarks, with total \textquotedblleft spin\textquotedblright $S$ and relative angular momentum $L_{12-34}$. Only the states for which $Q$ and $\bar Q$ have opposite charges are $C$ eigenvectors, with eigenvalues \cite{Estrada} $C=(-1)^{L_{12-34}+S}$. A discussion of G parity and its eigenvalues can be found in Ref.\cite{Santopinto:2006my}.
Tetraquark mesons do not have forbidden $J^{PC}$ combinations. 
Because of the Pauli principle, the tetraquark states must be antisymmetric for the exchange of the two quarks and the two antiquarks and it is, thus, necessary to study the permutational symmetry (i.e. the irreducible representations of the group $S_{2}$) of the color, flavor, spin and spatial parts of the wave functions of each subsystem. Moreover we have another constraint:   
only the singlet colour states are physical states. We have seen that there are two colour singlets allowed to the tetraquarks. It is better to write them by underlining their permutational $S_{2}$ symmetry, antisymmetric (A) or symmetric (S): $(qq)$ in $\;[\bar 3]_{C}\;(A)\;$ and $\;(\bar q\bar q)\;$ in $ [3]_{C}\; (A)$, or  $(qq)\;$ in $ [6]_{C} (S)$ and $\; (\bar q\bar q)\;$ in $\; [\bar 6]_{C}\; (S)$. 
Then we have to study the permutational symmetry of the spatial part of the two quarks (two antiquarks) states and the permutational symmetry of the SU(6)$_{sf}$ representations for a couple of quarks (antiquarks). The spatial, flavor, color and spin parts with given permutational symmetry ($S_{2}$) must then be combined together to obtain completely antisymmetric states under the exchange of the two quarks and the two antiquarks. The resulting states are listed in Table 
III of  Ref. \cite{Santopinto:2006my}.
In Table V, VI, VII and VIII of Ref. \cite{Santopinto:2006my} we write the possible flavor, spin and $J^{~PC}$ values for different orbital angular momenta. 

\section{\label{sec:spettrotetrascorrelato}The tetraquark spectrum}

In Ref.\cite{Iachello:1991fj} Iachello, Mukhopadhyay and Zhang developed a mass formula for $q\bar q$ mesons,  
\begin{equation}
M^{2}=(N_{n}M_{n}+N_{s}M_{s})^{2}+a \, \nu +b\, L+c\, S+d\, J+e\, M'^{2}_{iji'j'}+f\, M''^{2}_{iji'j'},
\label{eq:formulamassa}
\end{equation}
where $N_{n}$ is the non-strange quark and antiquark number, $M_{n}\equiv M_{u}=M_{d}$ is the non-strange constituent quark mass, $N_{s}$ is the strange quark and antiquark number, $M_{s}$ is the strange constituent quark mass, $\nu $ is the vibrational quantum number, $L$, $S$ and $J$ are the total orbital angular momentum, the total spin and the total angular momentum respectively, $M'^{2}_{iji'j'}$ and $M''^{2}_{iji'j'}$ are two phenomenological terms which act only on the lowest pseudoscalar mesons. The first acts only on the octect and encodes the unusually low masses of the eight Goldstone bosons, while the second acts on the $\eta $ and $\eta '$ mesons and encodes the non-negligible $q\bar q$ annihilation effecs that arise when the lowest mesons are flavour diagonal.
The flavor states are considered in the ideal mixing hypothesis, with the exception of the lowest pseudoscalar nonet whose mixing angle can be found in Ref.\cite{Iachello:1991fj}.
During the many years that have passed from the publication in 1991 of Iachello's article the values of the mesons masses reported by the PDG are changed in a considerable way. Thus, we have decided to update the fit of the Iachello model using the latest values reported by the PDG \cite{PDG} for the light $q \bar q$ mesons . The resulting parameters are reported in Ref. \cite{Santopinto:2006my}. 
\begin{table}[ph]
\begin{center}
\caption{The candidate tetraquark nonet. Experimental data and quantum numbers}
\begin{tabular}{l|l|l|l|l}
\hline \hline
Meson & $I^{G}(J^{PC})$ & $ N_{s}$ & $Mass \;(GeV)$ & Source \\
\hline
$a_{0}(980)$ & $1^{-}(0^{++})$ & 2 & $0.9847\pm 0.0012$  & PDG \cite{PDG} \\
$f_{0}(980)$ & $0^{+}(0^{++})$ & 2 & $0.980\pm 0.010$ &  PDG \cite{PDG}  \\
$f_{0}(600)$ & $0^{+}(0^{++})$ & 0 & $0.478\pm 0.024$ &  KLOE \cite{Aloisio:2002bt}  \\
$k(800)$ & $\frac{1}{2}(0^{+})$ & 1 & $0.797\pm 0.019$ &  E791 \cite{Aitala:2002kr}  \\
\hline \hline
\end{tabular}
\end{center}
\end{table}
The Iachello mass formula was developed for $q\bar q$ mesons. In order to describe uncorrelated tetraquark systems by means of an algebraic model one should use a new spectrum generating algebra for the spatial part, i.e. U(10) since we have nine spatial degrees of freedom. We have not addressed this difficult problem yet, but we chose to write the internal degrees of freedom part of the mass formula in the same way as it was done for the $q\bar q$ mesons. The splitting inside a given flavor multiplet, to which is also associated a given spin, can be described by the part of the mass formula that depends on the numbers of strange and non-strange quarks and antiquarks. 
Thus we can use, with the only purpose of determining the mass splitting of the candidate tetraquark nonet, see Ref.\cite{Santopinto:2006my},
\begin{equation}
\label{eq:tetrascorrelati}
M^{2}=const +(N_{n}M_{n}+N_{s}M_{s})^{2},
\end{equation}
where $const$ is a constant that includes all the spatial and spin dependence of the mass formula, and $M_{n}$ and $M_{s}$ are the masses of the constituent quarks as obtained from the previously discussed upgrade of the parameters of the Iachello mass formula.
We set the energy scale, i. e. we determine the constant $const$, by applying Eq.(\ref{eq:tetrascorrelati}) to the best-known candidate tetraquark, $a_{0}(980)$, see Ref.\cite{Santopinto:2006my}. Thus, the masses of the other mesons belonging to the same tetraquark nonet, predicted with our simple formula, are $M(\kappa(800))=0.726\;GeV$, $M(f_{0}(600))=0.354\;GeV$ and $M(f_{0}(980))=0.984\;GeV$ .
These values do not seem in very good agreement with the experimental values, even if, before reaching any conclusion, new experiments, especially on the poorly known $\kappa(800)$ and $f_{0}(600)$, are mandatory.

\section{\label{sec:diqantidiq}Diquark-antidiquark model}

 We think of the constituent diquark\footnote{For a discussion about the existence or not of the diquark degree of freedom and its importance in our model, please see Ref.\cite{Santopinto:2006my} and references therein.} as two correlated  constituent quarks with no internal spatial excitations, or at least we hypothesize that their internal spatial excitations will be higher in energy than the scale of masses of the resonances we will consider.  The tetraquark mesons are described in this model as composed of a constituent diquark, $(qq)$, and a constituent antidiquark, $(\bar q \bar q)$.
The diquark SU(3)$_{c}$ color representations are $[\bar 3]_{c}$ and $[6]_{c}$, while the antidiquark ones are $[3]_{c}$ and $[\bar 6]_{c}$, using the standard convention of denoting  color and flavor by the dimensions of their representation. As the tetraquark must be a color singlet, the possible diquark-antidiquark color combinations are  
\begin{subequations} 
\begin{eqnarray}
&  \text{diquark\; in}\; [\bar 3]_{c},\; \text{antidiquark\; in}\; [3]_{c} & \\
&   \text{diquark\; in}\; [6]_{c},\; \text{antidiquark\; in}\; [\bar 6]_{c} &
\end{eqnarray}
\end{subequations}
Diquarks (and antidiquarks) are made up of two identical fermions and so they have to satisfy the Pauli principle. Since we consider diquarks with no internal spatial excitations, their color-spin-flavor wave functions must be antisymmetric. This limits the possible representations to being only 
\begin{subequations}
\begin{eqnarray}
&  \text{color \; in} ~ [\bar 3]~ \text{(AS),\; spin-flavor\;in} [21]_{sf}~\text{(S)} & \\
&  \text{color \;in }[6]~\text{(S)} \text{,\; spin-flavor\; in~} [15]_{sf}~ 
\text{(AS)}  & 
\end{eqnarray}
\end{subequations}

The decomposition of these SU$_{sf}$(6) representations in terms of SU(3)$_{f}\otimes $ SU(2)$_{s}$ is 
 (in the notation $[\text{flavor\;repr.,\;spin}]$)
\begin{subequations}
\begin{eqnarray}
& [21]_{sf}=[\bar 3,0]\oplus [6,1] & \\
& [15]_{sf}=[\bar 3,1]\oplus [6,0] &
\end{eqnarray} 
\end{subequations}
Using the notation $|\text{flavor\;repr.,\;color\;repr.,\;spin}\rangle$, the diquark states corresponding  to color 
$[\bar 3]_{c}$ and  $[6]_{c}$ respectively, are
\begin{eqnarray}
&  |[\bar 3]_{f},[\bar 3]_{c},0\rangle, |[6]_{f},[\bar 3]_{c},1\rangle & \\
&  |[\bar 3]_{f},[6]_{c},1\rangle, |[6]_{f},[6]_{c},0\rangle  & 
\end{eqnarray} 
The antidiquark states are the conjugate of the above states.  

Following Refs.\cite{Jaffe:2004ph,Jaffe:1999ze} or Ref.\cite{Lichtenberg:1996fi},
we expect that color-sextet diquarks will be higher in energy than color-triplet diquarks or even that they will not be bound at all. Thus, we will consider only diquarks and antidiquarks in $[\bar 3]_{c}$ and $[3]_{c}$ color representations.

The tetraquark color-spin flavor states, obtained combining the allowed diquark and antidiquark states, are reported in Table XI of Ref.\cite{Santopinto:2006my}. Since diquarks are considered with no internal spatial excitations, the tetraquark states in this model are a subset of the tetraquark states previously derived. In particular they corresponds to the subset with 
$L_{13}=L_{24}=0$, and color $[\bar 3]_{c}\otimes [3]_{c}$. 
The relative orbital angular momentum between the diquark and the antidiquark is denoted by $L_{12-34}$; $S_{dq}$ and $ S_{d \bar q}$ are respectively the spin of the diquark and the spin of the antidiquark, and $S_{tot}$ is the total spin; $J$ is the total angular momentum.
 
Table XII of Ref.\cite{Santopinto:2006my} shows the corresponding flavor tetraquark states for each diquark and antidiquark content in the ideal mixing hypothesis. 

We have also determined the $J^{PC}$ quantum numbers of the tetraquarks in the diquark-antidiquark limit. 
We start from the possible quantum numbers classified for the uncorrelated tetraquark states
and then apply the restrictions for the diquark-antidiquark limit, $L_{13}=L_{24}=0$ and color 
$[\bar 3]_{c}\otimes [3]_{c}$. Thus, the parity of a tetraquark in the diquark-antidiquark limit is
$P=(-1)^{L_{12-34}}$,
while the charge conjugation (obviously only for its eigenstates) is
$C=(-1)^{L_{12-34}+S_{tot}}$.
In Ref.\cite{Santopinto:2006my} is also discussed the $G$ parity in the diquark-antidiquark limit.

The possible $J^{PC}$ combinations and diquark content of diquark-antidiquark systems with 
$L_{12-34}=0$ , $L_{12-34}=1$ and $L_{12-34}=2$ are reported in Ref.\cite{Santopinto:2006my}, in Tables XIII, XIV and XV respectively.

\subsection{\label{subsec:spettrodiqantiq}The tetraquark nonet spectrum in the diquark-antidiquark model.}
In the diquark-antidiquark limit we can use $U(4)\otimes SU(3)_{f}\otimes SU(2)_{s}\otimes SU(3)_{c}$ as spectrum generating algebra, by analogy with what was done by Iachello et al. in Ref.\cite{Iachello:1991re,Iachello:1991fj} for the normal mesons. The analogy between the tetraquark in the diquark-antidiquark limit and the $q\bar q$ mesons is even more evident if we consider that in a string model, as we can see in Refs.\cite{Johnson:1975sg,'tHooft:1974hx} the slopes of the Regge trajectories depend only on the color representation of the constituent particles.
Thus the slope of the trajectories of tetraquarks made up of a diquark in $[\bar 3]_{c}$ and an antidiquark in $[3]_{c}$ is the same as the slope of the trajectories of $q\bar q$ mesons.

Following all these considerations, it is evident that for the tetraquark in the diquark-antidiquark model we can use the same mass formula developed for the normal mesons, with the only difference that we have to replace the masses of the quark and the antiquark with those of the diquark and the antidiquark:
\begin{equation}
\label{eq:formulamassadqantidq}
M^{2}=(M_{qq}+M_{\bar q\bar q})^{2}+a\cdot n+b\cdot L_{12-34}+c\cdot S_{tot}+d\cdot J,
\end{equation}
where $M_{qq}$ and $M_{\bar q\bar q}$ are the diquark and antidiquark masses, $n$ is a vibrational quantum number, $L_{12-34}$ the relative orbital angular momentum, $S_{tot}$ the total spin and $J$ the total angular momentum.

The diquark masses are unknown parameters and are determined by fitting the mass formula Eq.(\ref{eq:formulamassadqantidq}) with the mass values of the tetraquark candidate nonet\footnote{This nonet has quantum numbers $n=L_{12-34}=S_{tot}=J=0$, so we do not need to know the parameters $a$, $b$, $c$ and $d$.} $a_{0}(980)$, $f_{0}(980)$, $f_{0}(600)$ and $\kappa (800)$. 
We consider the candidate tetraquark nonet as the fundamental tetraquark multiplet and so it contains the lighter diquarks, i.e. scalar diquarks. 
\begin{table}
\begin{center}
\caption{\label{tab:nonettotetra2diq} Quantum numbers of the candidate tetraquark nonet. $\kappa (800)$ corresponds to $[n,n]\overline{[n,s]}$ and also to its conjugate.}
\begin{tabular}{|c|c|c|c|c|}
\hline
Meson & Mass ($GeV$)& Diquark content & $I^{G}(J^{PC})$ & Source \\
\hline
$a_{0}(980)$ & $0.9847\pm 0.0012$ & $[n,s]\overline{[n,s]}$ & $1^{-}(0^{++})$ & PDG \cite{PDG} \\
\hline
$f_{0}(980)$ & $0.980\pm 0.010$ & $[n,s]\overline{[n,s]}$ & $0^{+}(0^{++})$ &  PDG \cite{PDG}  \\
\hline
$f_{0}(600)$ & $0.478\pm 0.024$ & $[n,n]\overline{[n,n]}$ & $0^{+}(0^{++})$ &  KLOE \cite{Aloisio:2002bt}  \\
\hline
$\kappa (800)$ & $0.797\pm 0.019$ & $[n,n]\overline{[n,s]}$ & $\frac{1}{2}(0^{+})$ & E791 \cite{Aitala:2002kr}  \\
\hline
\end{tabular}
\end{center}
\end{table}

The masses of the scalar diquarks resulting from the fit are:
\begin{equation}
\label{eq:massadiquarkscalari}
M_{[n,n]}  =  0.275\;GeV,\;\;M_{[n,s]}  =  0.492\;GeV  
\end{equation}
From the fit we obtain also the following masses of the candidate tetraquark nonet:
\begin{subequations}
\begin{eqnarray}
 & M_{a_{0}(980)}=M_{f_{0}(980)}= 0.984\;GeV & \\
 & M_{f_{0}(600)}= 0.550\;GeV & \\
 & M_{\kappa (800)}= 0.767\;GeV. & 
\end{eqnarray}
\end{subequations}
\begin{figure}[h]
\begin{center}
\includegraphics[height=6 cm]{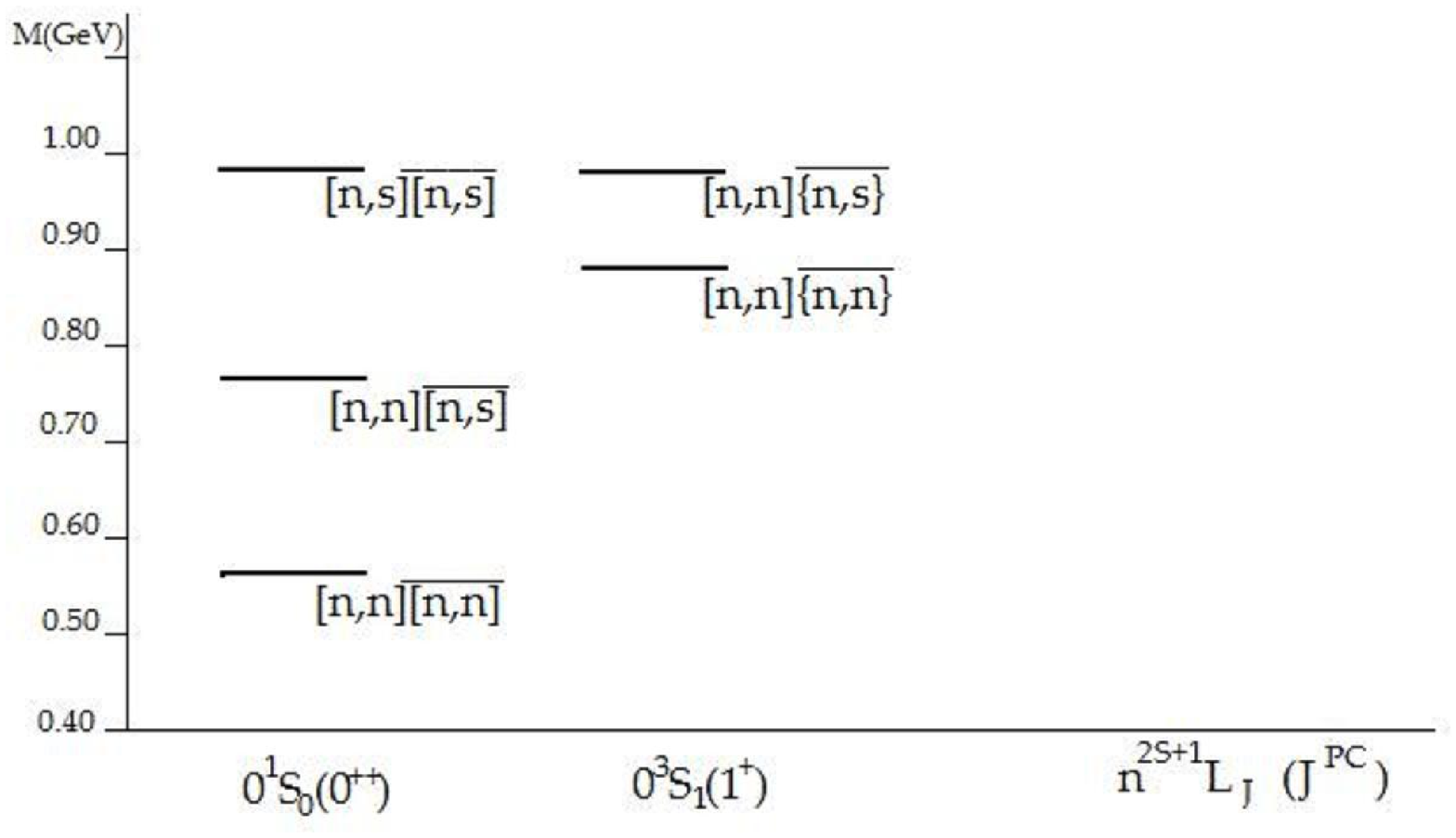}
\caption{Lowest part (below $1\;GeV$) of the tetraquark spectrum in the diquark-antidiquark model.} \label{Fig:spettrodiqadiq}
\end{center}
\end{figure}
These masses are much closer to the experimental values reported in Table \ref{tab:nonettotetra2diq} than the masses obtained in the uncorrelated tetraquark model of Section \ref{sec:spettrotetrascorrelato}. However we again underline that, before reaching any conclusion, new experiments are necessary
also to be sure of the existence of all the states of the scalar nonet. If the existence of only some states of the nonet will be confirmed a different kind of clusterization will emerge, and we have still not studied  this limit in the algebraic framework. Moreover, we have still not studied the decays of these states and the study of their decay properties can give a better insight into their nature.

The mass formula Eq. \ref{eq:formulamassadqantidq} can be used to predict all the spectrum of the tetraquarks in the diquark-antidiquark model.  
In Fig. \ref{Fig:spettrodiqadiq} we include as a preview of this work (still in progress) the lowest (below $1\;GeV$) part of this spectrum.


\printindex{author}{Author Index}
\blankpage

\printindex{subject}{Subject Index}
 \blankpage\end{document}